\documentclass[%
 reprint,
 amsmath,amssymb,
 aps,
]{revtex4-1}

\usepackage{graphicx}% Include figure files
\usepackage{dcolumn}% Align table columns on decimal point
\usepackage{bm}% bold math
\usepackage[T1]{fontenc}
\usepackage[latin9]{inputenc}
\usepackage{amssymb}
\usepackage{color}
\usepackage{graphicx,color}
\usepackage{amsmath, amssymb, graphics}
\usepackage{qcircuit}
\usepackage{pgf, tikz}
\usetikzlibrary{arrows, automata}
\makeatletter
\numberwithin{equation}{section} %% Comment out for sequentially-numbered
\numberwithin{figure}{section} %% Comment out for sequentially-numbered
  \@ifundefined{theoremstyle}{\usepackage{amsthm}}{}
  \theoremstyle{plain}
  
  \theoremstyle{plain}
  
  \theoremstyle{plain}
  
  \theoremstyle{remark}
  
  \theoremstyle{remark}
  
  \theoremstyle{plain}

\tikzset{
circleA/.style={
  circle,
  inner sep=0pt,
  text width=6mm,
  align=center,
  draw=black,
  fill=white
  }
}

\smallskip
\def\<{{\langle }}
\def\>{{\rangle }}

\def\ket#1{|#1\rangle}

\smallskip
\def\<{{\langle }}
\def\>{{\rangle }}

\begin{document}

%\preprint{APS/123-QED}

\title{Four qubits generated by Clifford gates}% Force line breaks with \\

\author{Oscar Perdomo}
 \email{perdomoosm@ccsu.edu}
%\altaffiliation[Also at ]{Central Connecticut State University}%Lines break automatically or can be forced with \\
\author{Fr\'ed\'eric Latour}%
 \email{latourfre@ccsu.edu}
\affiliation{
Central Connecticut State University 
}
%



\date{\today}% It is always \today, today,

\begin{abstract}  The Clifford group is the set of gates generated by CNOT gates and the two local gates $P=\begin{pmatrix} 1&0\\0&i\end{pmatrix}$ and  $H=\frac{1}{\sqrt{2}} \begin{pmatrix} 1&1\\1&-1\end{pmatrix}$. We will say that a $n$-qubit state is a Clifford state if it can be prepared using Clifford gates, this is, $\ket{\phi}$ is Clifford if 
$\ket{\phi}=U\ket{0\dots 0}$ where $U$ is a Clifford gate. In this paper we study the set of all $4$-qubit Clifford states.  We prove that there are $293760$ states and their entanglement entropy must be either $0$, $2/3$, $1$, $4/3$ and $5/3$. We also show that any pair of these states can be connected using local gates and at most $3$ CNOT gates. We also study the Clifford states with real entries under the action of the subgroup $\cal{C}\mathbb{R}$ of  Clifford gates with real entries. This time we show that every pair of Clifford states with real entries can be connected with at most 5 CNOT gates and local gates in $\cal{C}\mathbb{R}$. Finally we show that the set of populations of these states include all 16 possibilities of the form $\ket{i_1i_2i_3i_4}$, it includes all 120 possibilities of the form $\frac{1}{\sqrt{2}}\ket{i_1i_2i_3i_4}+\frac{1}{\sqrt{2}}\ket{j_1j_2j_3j_4}$, in other words, it includes all 120 possibilities where only two elements in the canonical basis show up with equal probability $\frac{1}{2}$. It include only 140 possibilities where exactly four elements in the canonical basis shows up with equal probability $\frac{1}{4}$. It include only 30 possibilities where exactly 8 elements in the canonical basis shows up with equal probability and finally it include the case where all the 16 elements in the canonical base show up with equal probability. Overall, only 307 possible populations are possible. The link   \url{https://youtu.be/42MI6ks2_eU} leads you to a YouTube video that explains the most important results in this paper.
\end{abstract}
%\begin{description}
%\end{description}

\maketitle

%\tableofcontents

%$\ket{\phi_1}$ and $\ket{\phi_2}$,

\section{Introduction} Given two $n$ qubit pure states, an interesting and difficult question in quantum computing is the one of finding an ``efficient'' circuit that connects these two states. By efficient we mean that the number of non local gates is minimum. In this paper we answer this question for the collection of $4$-qubit Clifford states under the assumption that we can only use gates from the Clifford group. The way we study these states is by identify them using the equivalence relation: $\ket{\phi_1}\sim\ket{\phi_2}$ anytime $\ket{\phi_1}=U\ket{\phi_2}$ where $U$ is a local Clifford gate. The quotient space of this relation has 18 orbits, namely:

\begin{eqnarray*} \Lambda&=&\{S_0, T_{12}, T_{13}, T_{14}, T_{23}, T_{24}, T_{34}, U_{\frac{12}{34}}, U_{\frac{13}{24}}, U_{\frac{14}{23}}, \\ & & V_1, V_2, V_3, V_4, W, X_{\frac{12}{34}}, X_{\frac{13}{24}}, X_{\frac{14}{23}} \}
\end{eqnarray*}

While the above notation may seem arbitrary, it has the advantage of making clear the action of the Controlled NOT gates on each equivalence class, as we will see. Also, in the above notation, a subscript consisting of two numbers is treated as an unordered pair; for instance, $T_{31} = T_{13}.$ Also, a subscript consisting of two pairs separated by a horizontal bar or slash is treated as an unordered pair of unordered pairs; for instance, $U_{43/12} = U_{\frac{12}{34}}.$ We will use the overline $\overline{S}$ to denote the complement of $S$ with respect to $1234;$ for instance, $V_{\overline{124}} = V_3.$

We also consider the set of all possible states that can be generated by the set $\cal{C}\mathbb{R}$ of  Clifford gates with real entries. These gates can be viewed as those generates by CNOT gates, the Hadamard gate $H$ and the $X$ gate $\begin{pmatrix} 1&0\\0&-1\end{pmatrix}$. We prove that  8640 states are generated with gates in $\cal{C}\mathbb{R}$  and when we identify them to be equivalent if there a local in $\cal{C}\mathbb{R}$  that send one state into the other, we obtain 29 orbits. We will explain how to navigate these orbits using CNOT gates.

%A_{74}^0,B_{83}^{2/3},C_{83}^{2/3},D_{83}^{2/3},E_{83}^{2/3},F_{83}^{2/3}, G_{83}^{2/3}, H_{84}^1, %I_{92}^{4/3},\\ & & J_{84}^1,
% K_{92}^{4/3}, L_{84}^1,M_{92}^{4/3}, N_{84}^{1}, O_{84}^{1},  P_{94}^{5/3}, Q_{94}^{5/3}, R_{94}^{5/3} \} 

%The superscript in the name of each set indicates the entanglement entropy of the elements in the given %orbit. The subscript indicates the number of elements in the orbit. If the subscript is $ij$ then the number %of elements is $2^i3^j.$

\section{Main results for Clifford states}
The following table gives the size of each equivalence class, as well as the entanglement entropy of its elements:

\begin{center}
\setlength\extrarowheight{1mm}
\begin{tabular}{|c|c|c|} %the | are vertical bars, not lowercase Ls
\hline
class type & size & entropy \\
\hline
$S_0$ & $2^7 3^4 = 10368$ & $0$ \\
\hline 
$T_{ab}$ & $2^8 3^3 = 6912$ & $\frac{2}{3}$ \\
\hline
$U_{ab/cd}$ & $2^9 3^2 = 4608$ & $\frac{4}{3}$ \\
\hline 
$V_a$ & $2^8 3^4 = 20736$ & $1$ \\
\hline 
$W$ & $2^8 3^4 = 20736$ & $1$ \\
\hline 
$X_{ab/cd}$ & $2^9 3^4 =41472$ & $\frac{5}{3}$ \\
\hline
\end{tabular}
\end{center}

Adding up the sizes of all 18 equivalence classes, we get $10368 + 6 \times 6912 + 3 \times 4608 + 5 \times 20736 + 3 \times 41472 = 293760,$ which is therefore the total number of states.

The following diagram shows the connectivity with Controlled NOT gates between the orbits:

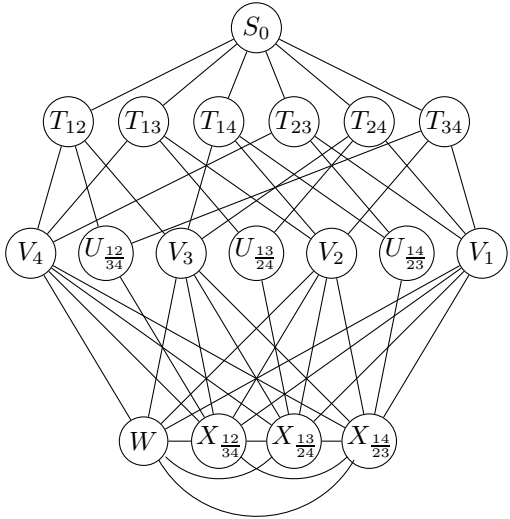
\begin{center}
    \begin{tikzpicture}
        \node at (0,0) [circleA] (A) {$S_0$};
        \node at (-0.5,-1.25) [circleA] (D) {$T_{14}$};
        \node at (-1.5,-1.25) [circleA] (C) {$T_{13}$};
        \node at (-2.5,-1.25) [circleA] (B) {$T_{12}$};
        \node at (0.5,-1.25) [circleA] (E) {$T_{23}$};
        \node at (1.5,-1.25) [circleA] (F) {$T_{24}$};
        \node at (2.5,-1.25) [circleA] (G) {$T_{34}$};
        \node at (0,-3) [circleA] (K) {$U_{\frac{13}{24}}$};
        \node at (-1,-3) [circleA] (J) {$V_3$};
        \node at (-2,-3) [circleA] (I) {$U_{\frac{12}{34}}$};
        \node at (-3,-3) [circleA] (H) {$V_4$};
        \node at (1,-3) [circleA] (L) {$V_2$};
        \node at (2,-3.) [circleA] (M) {$U_{\frac{14}{23}}$};
        \node at (3,-3) [circleA] (N) {$V_1$};
        \node at (-1.5,-5.5) [circleA] (O) {$W$};
        \node at (-0.5,-5.5) [circleA] (P) {$X_{\frac{12}{34}}$};
        \node at (0.5,-5.5) [circleA] (Q) {$X_{\frac{13}{24}}$};
        \node at (1.5,-5.5) [circleA] (R) {$X_{\frac{14}{23}}$};
        \draw (A) -- (B); 
        \draw (A) -- (C);
        \draw (A) -- (D);
        \draw (A) -- (E); 
        \draw (A) -- (F);
        \draw (A) -- (G);
        \draw (B) -- (H); 
        \draw (B) -- (I);
        \draw (B) -- (J);
        \draw (C) -- (H); 
        \draw (C) -- (K);
        \draw (C) -- (L);
        \draw (D) -- (J); 
        \draw (D) -- (L);
        \draw (D) -- (M);
        \draw (E) -- (H); 
        \draw (E) -- (M);
        \draw (E) -- (N);
        \draw (F) -- (J); 
        \draw (F) -- (K);
        \draw (F) -- (N);
        \draw (G) -- (I); 
        \draw (G) -- (L);
        \draw (G) -- (N);
        \draw (O) -- (H); 
        \draw (O) -- (J); 
        \draw (O) -- (L);
        \draw (O) -- (N);
        \draw (P) -- (H); 
        \draw (P) -- (J); 
        \draw (P) -- (L);
        \draw (P) -- (N);
        \draw (Q) -- (H); 
        \draw (Q) -- (J); 
        \draw (Q) -- (L);
        \draw (Q) -- (N);
        \draw (R) -- (H); 
        \draw (R) -- (J); 
        \draw (R) -- (L);
        \draw (R) -- (N);
        \draw (I) -- (P);
        \draw (K) -- (Q);
        \draw (M) -- (R);
        \draw (P) -- (Q);
        \draw (Q) -- (R);
        \draw (O) -- (P);
        \draw (0.5,-5) ++( 225 : 1 ) arc ( 225:315:1 );
        \draw (-0.5,-5) ++( 225 : 1 ) arc ( 225:315:1 );
        \draw (0,-5) ++( 210 : 1.5 ) arc ( 210:330:1.5 );
    \end{tikzpicture}
    \end{center}
    
The results of applying each Controlled NOT gate on the elements of each equivalence class are given by the following table:

\begin{center}
\setlength\extrarowheight{1mm}
\begin{tabular}{|c|c|c|} %the | are vertical bars, not lowercase Ls
\hline
class type & controlled NOT gate & number of elements \\
& used: $CZ(i,j)$ & mapped to each class \\
\hline
$S_0$ & all & 5760 to $S_0$ \\
& & 4608 to $T_{ij}$ \\
\hline 
$T_{ab}$ & $ij=ab$ & 2304 to $T_{ab}$ \\
& & 4608 to $S_0$ \\
\hline 
$T_{ab}$ & $ij=ac,bc$ & 2304 to $T_{ab}$ \\
& but $ij \neq ab$ & 4608 to $V_{\overline{abc}}$ \\
\hline 
$T_{ab}$ & $ij=\overline{ab}$ & 3840 to $T_{ab}$ \\
& & 3072 to $U_{ab/\overline{ab}}$ \\
\hline
$U_{ab/\overline{ab}}$ & $ij = ab, \overline{ab}$ & 3072 to $T_{\overline{ij}}$ \\
& & 1526 to $U_{ab/\overline{ab}}$ \\
\hline
$U_{ab/\overline{ab}}$ & $ij\neq ab, \overline{ab}$ & 4608 to $X_{ab/\overline{ab}}$ \\
\hline 
$V_a$ & $ij=ac$ & 6912 to $V_a$ \\
& & 4608 to $W$ \\
& & 9216 to $X_{ac/\overline{ac}}$ \\
\hline 
$V_a$ & $i \neq a, j \neq a$ & 11520 to $V_a$ \\
& & 4608 to $T_{\overline{ai}}$ \\
& & 4608 to $T_{\overline{aj}}$ \\
\hline 
$W$ & all & 2304 to $W$ \\
& & 9216 to $X_{ij/\overline{ij}}$ \\
& & 4608 each to $V_i, V_j$ \\
\hline 
$X_{{ab}/{\overline{ab}}}$ & $ij = ab, \overline{ab}$ & 13824 to $X_{{ab}/{\overline{ab}}}$ \\
& & 9216 each to $V_i, V_j, W$ \\
\hline 
$X_{{ab}/{\overline{ab}}}$ & $ij \neq ab, \overline{ab}$ & 18432 to each $X,$ \\
& & other than $X_{ij/\overline{ij}}$ \\
& & 4608 to $U_{ab/\overline{ab}}$ \\
\hline
\end{tabular}
\end{center}

\section{Main results for Clifford states with real entries}
    
The quotient space of this relation has 29 orbits, namely:

\begin{eqnarray*} \Lambda&=&\{S_0^r, T_{12}^r, T_{13}^r, T_{14}^r, T_{23}^r, T_{24}^r, T_{34}^r, U_{\frac{12}{34}}^r, U_{\frac{13}{24}}^r, U_{\frac{14}{23}}^r, \\ & & V_1^r, V_2^r, V_3^r, V_4^r, \hat{V}_1^r, \hat{V}_2^r, \hat{V}_3^r, \hat{V}_4^r, W^r, \hat{W}^r,  \\ & & X_{\frac{12}{34}}^r, X_{\frac{13}{24}}^r, X_{\frac{14}{23}}^r, \hat{X}_{12}^r, \hat{X}_{13}^r, \hat{X}_{14}^r, \hat{X}_{23}^r, \hat{X}_{24}^r, \hat{X}_{34}^r\}
\end{eqnarray*}

The following table gives the size of each equivalence class, as well as the entanglement entropy of its elements:

\begin{center}
\setlength\extrarowheight{1mm}
\begin{tabular}{|c|c|c|} %the | are vertical bars, not lowercase Ls
\hline
class type & size & entropy \\
\hline
$S_0^r$ & $2^9 = 512$ & $0$ \\
\hline 
$T_{ab}^r$ & $2^8  = 256$ & $\frac{2}{3}$ \\
\hline
$U_{ab/cd}^r$ & $2^7 = 128$ & $\frac{4}{3}$ \\
\hline 
$V_a^r$ & $2^9 = 512$ & $1$ \\
\hline
$\hat{V}_a^r$ & $2^7 = 128$ & $1$ \\
\hline 
$W^r$ & $2^9 = 512$ & $1$ \\
\hline 
$\hat{W}^r$ & $2^6 = 64$ & $1$ \\
\hline 
$X_{ab/cd}^r$ & $2^9 = 512$ & $\frac{5}{3}$ \\
\hline
$\hat{X}_{ab}^r$ & $2^8 = 256$ & $\frac{5}{3}$ \\
\hline
\end{tabular}
\end{center}

Adding up the sizes of all 18 equivalence classes, we get $9 \times 512 + 12 \times 256 +  7 \times 128 +  64  = 8640,$ which is therefore the total number of states.

The following diagram shows the connectivity with Controlled NOT gates between the orbits:

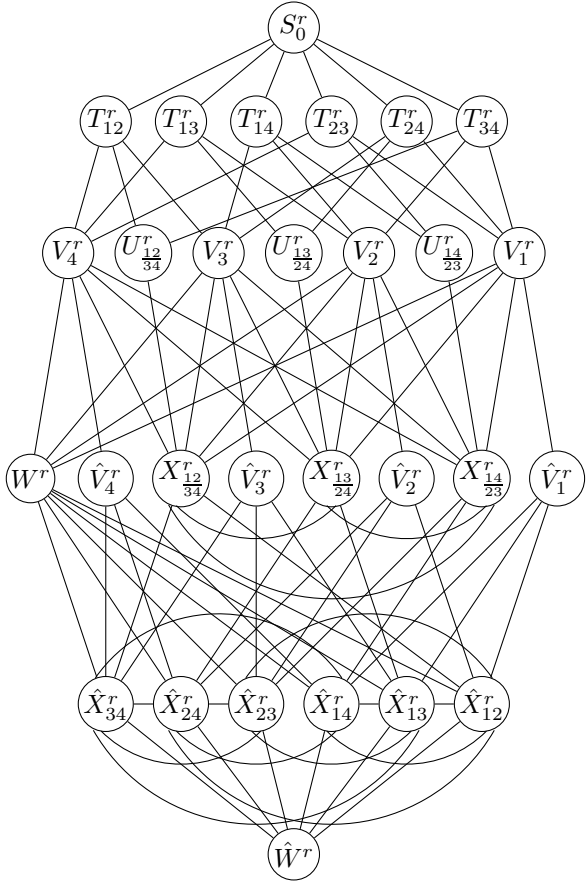
\begin{center}
    \begin{tikzpicture}
        \node at (0,0) [circleA] (A) {$S_0^r$};
        \node at (-0.5,-1.25) [circleA] (D) {$T_{14}^r$};
        \node at (-1.5,-1.25) [circleA] (C) {$T_{13}^r$};
        \node at (-2.5,-1.25) [circleA] (B) {$T_{12}^r$};
        \node at (0.5,-1.25) [circleA] (E) {$T_{23}^r$};
        \node at (1.5,-1.25) [circleA] (F) {$T_{24}^r$};
        \node at (2.5,-1.25) [circleA] (G) {$T_{34}^r$};
        \node at (0,-3) [circleA] (K) {$U_{\frac{13}{24}}^r$};
        \node at (-1,-3) [circleA] (J) {$V_3^r$};
        \node at (-2,-3) [circleA] (I) {$U_{\frac{12}{34}}^r$};
        \node at (-3,-3) [circleA] (H) {$V_4^r$};
        \node at (1,-3) [circleA] (L) {$V_2^r$};
        \node at (2,-3.) [circleA] (M) {$U_{\frac{14}{23}}^r$};
        \node at (3,-3) [circleA] (N) {$V_1^r$};
        \node at (-3.5,-6) [circleA] (O) {$W^r$};
        \node at (-1.5,-6) [circleA] (P) {$X_{\frac{12}{34}}^r$};
        \node at (0.5,-6) [circleA] (Q) {$X_{\frac{13}{24}}^r$};
        \node at (2.5,-6) [circleA] (R) {$X_{\frac{14}{23}}^r$};
        \node at (-2.5,-6) [circleA] (mH) {$\hat{V}_4^r$};
        \node at (-0.5,-6) [circleA] (mJ) {$\hat{V}_3^r$};
        \node at (1.5,-6) [circleA] (mL) {$\hat{V}_2^r$};
        \node at (3.5,-6) [circleA] (mN) {$\hat{V}_1^r$};
        \node at (0.5,-9) [circleA] (mD) {$\hat{X}_{14}^r$};
        \node at (1.5,-9) [circleA] (mC) {$\hat{X}_{13}^r$};
        \node at (2.5,-9) [circleA] (mB) {$\hat{X}_{12}^r$};
        \node at (-0.5,-9) [circleA] (mE) {$\hat{X}_{23}^r$};
        \node at (-1.5,-9) [circleA] (mF) {$\hat{X}_{24}^r$};
        \node at (-2.5,-9) [circleA] (mG) {$\hat{X}_{34}^r$};
        \node at (0,-11) [circleA] (mO) {$\hat{W}^r$};
        \draw (A) -- (B); 
        \draw (A) -- (C);
        \draw (A) -- (D);
        \draw (A) -- (E); 
        \draw (A) -- (F);
        \draw (A) -- (G);
        \draw (B) -- (H); 
        \draw (B) -- (I);
        \draw (B) -- (J);
        \draw (C) -- (H); 
        \draw (C) -- (K);
        \draw (C) -- (L);
        \draw (D) -- (J); 
        \draw (D) -- (L);
        \draw (D) -- (M);
        \draw (E) -- (H); 
        \draw (E) -- (M);
        \draw (E) -- (N);
        \draw (F) -- (J); 
        \draw (F) -- (K);
        \draw (F) -- (N);
        \draw (G) -- (I); 
        \draw (G) -- (L);
        \draw (G) -- (N);
        \draw (O) -- (H); 
        \draw (O) -- (J); 
        \draw (O) -- (L);
        \draw (O) -- (N);
        \draw (P) -- (H); 
        \draw (P) -- (J); 
        \draw (P) -- (L);
        \draw (P) -- (N);
        \draw (Q) -- (H); 
        \draw (Q) -- (J); 
        \draw (Q) -- (L);
        \draw (Q) -- (N);
        \draw (R) -- (H); 
        \draw (R) -- (J); 
        \draw (R) -- (L);
        \draw (R) -- (N);
        \draw (I) -- (P);
        \draw (K) -- (Q);
        \draw (M) -- (R);
        \draw (H) -- (mH);
        \draw (J) -- (mJ);
        \draw (L) -- (mL);
        \draw (N) -- (mN);
        \draw (mB) -- (mO);
        \draw (mC) -- (mO);
        \draw (mD) -- (mO);
        \draw (mE) -- (mO);
        \draw (mF) -- (mO);
        \draw (mG) -- (mO);
        \draw (mB) -- (O);
        \draw (mC) -- (O);
        \draw (mD) -- (O);
        \draw (mE) -- (O);
        \draw (mF) -- (O);
        \draw (mG) -- (O);
        \draw (mB) -- (P);
        \draw (mC) -- (Q);
        \draw (mD) -- (R);
        \draw (mE) -- (R);
        \draw (mF) -- (Q);
        \draw (mG) -- (P);
        \draw (mD) -- (mH);
        \draw (mF) -- (mH);
        \draw (mG) -- (mH);
        \draw (mC) -- (mJ);
        \draw (mE) -- (mJ);
        \draw (mG) -- (mJ);
        \draw (mB) -- (mL);
        \draw (mE) -- (mL);
        \draw (mF) -- (mL);
        \draw (mB) -- (mN);
        \draw (mC) -- (mN);
        \draw (mD) -- (mN);
        \draw (mF) -- (mG);
        \draw (mE) -- (mF);
        \draw (mC) -- (mD);
        \draw (mB) -- (mC);
        \draw (1.5,-5.3) ++( 225 : 1.5) arc ( 225:315:1.5 );
        \draw (-0.5,-5.3) ++( 225 : 1.5 ) arc ( 225:315:1.5 );
        \draw (0.5,-5.1) ++( 210 : 2.5 ) arc ( 210:330:2.5 );
        \draw (1.5,-8.3) ++( 225 : 1.5) arc ( 225:315:1.5 );
        \draw (-0.5,-8.3) ++( 225 : 1.5 ) arc ( 225:315:1.5 );
        \draw (0.5,-8.3) ++( 225 : 1.5) arc ( 225:315:1.5 );
        \draw (-1.5,-8.3) ++( 225 : 1.5 ) arc ( 225:315:1.5 );
        \draw (0.5,-8.1) ++( 210 : 2.5) arc ( 210:330:2.5 );
        \draw (-0.5,-8.1) ++( 210 : 2.5) arc ( 210:330:2.5 );
        \draw (1,-9.8) ++( 35 : 2) arc ( 35:145:2 );
        \draw (-1,-9.8) ++( 35 : 2 ) arc ( 35:145:2 );
    \end{tikzpicture}
    \end{center}

%        \draw (0.5,-5) ++( 225 : 1 ) arc ( 225:315:1 );
 %       \draw (-0.5,-5) ++( 225 : 1 ) arc ( 225:315:1 );
  %      \draw (0,-5) ++( 210 : 1.5 ) arc ( 210:330:1.5 );  
    
The results of applying each Controlled NOT gate on the elements of each equivalence class are given by the following table:

\begin{center}
\setlength\extrarowheight{1mm}
\begin{tabular}{|c|c|c|} %the | are vertical bars, not lowercase Ls
\hline
class type & controlled NOT gate & number of elements \\
& used: $CZ(i,j)$ & mapped to each class \\
\hline
$S_0^r$ & all & 384 to $S_0^r$ \\
& & 128 to $T_{ij}^r$ \\
\hline 
$T_{ab}^r$ & $ij=ab$ &128 each to $T_{ab}^r, S_0^r$ \\
\hline 
$T_{ab}^r$ & $ij=ac,bc$ & 128 each to $T_{ab}^r, V_{\overline{abc}}^r$ \\
& but $ij \neq ab$ &  \\
\hline 
$T_{ab}^r$ & $ij=\overline{ab}$ & 192 to $T_{ab}^r$ \\
& & 64 to $U_{ab/\overline{ab}}^r$ \\
\hline
$U_{ab/\overline{ab}}^r$ & $ij = ab, \overline{ab}$ & 64 each to $T_{\overline{ab}}^r, U_{ab/\overline{ab}}^r$ \\
\hline
$U_{ab/\overline{ab}}^r$ & $ij\neq ab, \overline{ab}$ & 128 to $X_{ab/\overline{ab}}^r$ \\
\hline 
$V_a^r$ & $ij=ac$ & 256 to $V_a^r$ \\
& & 128 each to $W^r, X_{ac/\overline{ac}}^r$ \\
\hline 
$V_a^r$ & $i \neq a, j \neq a$ & 128 each to \\
& & $V_a^r, T_{\overline{ai}}^r, T_{\overline{aj}}^r, \hat{V}_a^r$ \\
\hline 
$\hat{V}_a^r$ & $ij=ac$ & 64 each to $\hat{V}_a^r, \hat{X}_{ac}^r$ \\
\hline 
$\hat{V}_a^r$ & $i \neq a, j \neq a$ & 128 to $V_a^r$ \\
\hline 
$W$ & all & 128 each to  \\
& & $W^r, \hat{X}_{\overline{ij}}^r, V_i^r, V_j^r$ \\
\hline 
$\hat{W}$ & all & 64 to $\hat{X}_{\overline{ij}}^r$ \\
\hline 
$X_{{ab}/{\overline{ab}}}^r$ & $ij = ab, \overline{ab}$ & 128 each to \\
& & $V_i^r, V_j^r, \hat{X}_{\overline{ij}}^r, X_{{ab}/{\overline{ab}}}^r$ \\
\hline 
$X_{{ab}/{\overline{ab}}}^r$ & $ij = ac, c\neq b$ & 128 each to \\
& & $U_{{ab}/{\overline{ab}}}^r, X_{{bc}/{\overline{bc}}}^r, \hat{X}_{ab}^r, \hat{X}_{\overline{ab}}^r$ \\
\hline 
$\hat{X}_{ab}^r$ & $ij=ab$ &64 each to \\
& & $\hat{W}^r, \hat{X}_{ab}^r, \hat{V}_a^r, \hat{V}_b^r$ \\
\hline 
$\hat{X}_{ab}^r$ & $ij=ac,bc$ &  128 each to \\
& but $ij \neq ab$ &  $\hat{X}_{\overline{bc}}^r, X_{{ab}/{\overline{ab}}}^r$ \\
\hline 
$\hat{X}_{ab}^r$ & $ij=\overline{ab}$ & 128 each to $W^r, X_{{ab}/{\overline{ab}}}^r$ \\
\hline
\end{tabular}
\end{center}

\section{Conclusion}
\begin{enumerate}
\item
This paper can be viewed as a continuation of the paper \cite{P}. One of the main differences is that the local group of Clifford gates for 3 qubits  is considerable smaller than its corresponding group for 4 qubits. For this reason, while the orbits for the 3 qubits Clifford states were obtained by directly multiplying by the  group of local gates, for the case for 4 qubits we have used the following procedure: (i)  We took a set of gates $G$ that  generates all local Clifford gates. (ii) Given the state $\ket{\phi_0}$, we computed the set $\Lambda_1=\{g\ket{\phi_0}\, :\ g\in G\}$, $\Lambda_2=\{g\ket{\phi}\, :\ g\in G\, ,\, \ket{\phi}\in \Lambda_1 \}$, $\dots$ until $\Lambda_{i+1}=\Lambda_i$. It is not difficult to show that $\Lambda_i$ is the orbit that contains the state $\ket{\phi_0}$.

\item

It is known that there are not 4 qubit states that are absolutely maximal entangled (AME), this is, it is impossible that have a 4 qubit state with reduced density matrices of all six pairs of qubits being a multiple of the identity matrix \cite{H}. An AME 4 qubit state would have entanglement entropy equal to 2 and as we have just mentioned, this value is not reached. So far, the maximum known value for the entanglement entropy is $\ln(12)/\ln(4)\approx 1.792\dots$. The maximum value for the entanglement entropy among the Clifford states is $5/3\approx 1.666\dots$. Moreover the only possible values for the entanglement entropy defined on the Clifford 4 qubit states are 0, $2/3$, $1$, $4/3$ and $5/3$.
\item

Since there are 29 orbits when we consider the Clifford states with real amplitudes and only 18 orbits for the Clifford states, we conclude that there exist pair of 4 qubit states with real amplitudes that can be connected with a Clifford gate but not with a Clifford gate with real entries. These pair of 4 qubits do not exist among the orbits $S_0$, $T_{ab}$, $U_{ab/\overline{ab}}$. They do exists for the rest of the orbits and they are responsible for the splitting of states with real amplitudes in $W$ into the orbits the sets $W^r$ and $\hat{W}^r$, those in $X_{ab/\overline{ab}}$ into the sets $X^r_{ab/\overline{ab}}$, $X^r_{ab}$ and $X^r_{\overline{ab}}$, and  those in $V_{a}$ into $V^r_a$ and $\hat{V}^r_a$.

\item

Any pair of Clifford states can be connected using local gates and at most 3 CNOT gates. Any pair of Clifford state with real amplitudes can be connected using local Clifford gates with real entries and at most 5 CNOT gates.

\item

The population distribution in each Clifford is uniform. Moreover the values of the probabilities are $\{1\}$, $\{1/2,1/2\}$ or $\{1/4.1/4,1/4,1/4\}$ or $\{1/8.\dots,1/8\}$ or $\{1/16.\dots,1/16\}$.

\item
Let us denote by $\ket{\hbox{\bf 0}}=\ket{0000}$, $\ket{\hbox{\bf 1}}=\ket{0001}$, $\dots$, $\ket{\hbox{\bf 15}}=\ket{1111}$.  Even though the set of Clifford states contains all the  populations of the form $\ket{\hbox{\bf i}}$ with $i=0,\dots, 15$ and all the possible populations of the form 
$\frac{1}{\sqrt{2}} (\ket{\hbox{\bf i}}+\ket{\hbox{\bf i}})$ with $i\ne j$, we have that  not every state of the form $\frac{1}{2} (\ket{\hbox{\bf i}}+\ket{\hbox{\bf i}}+\ket{\hbox{\bf k}}+\ket{\hbox{\bf l}})$ is reached. Neither are all the possible combinations of those with $\frac{1}{8}$ uniform probability or all the possible combination with  $\frac{1}{16}$ uniform probability.

\end{enumerate}

\section{Apendix}

In order to describe each one of the orbits we only need to display one qubit state in each orbit. The following table presents these states.

\begin{center}
\setlength\extrarowheight{1mm}
\begin{tabular}{|c|c|c|} %the | are vertical bars, not lowercase Ls
\hline
class type &  a qubit state in it \\
\hline
$S_0, \, S_0^r$ &  $\ket{0000}$ \\
\hline 
$T_{12},\, T_{12}^r$ & $\frac{1}{\sqrt{2}}(\ket{1110}-\ket{1101})$ \\
\hline 
$T_{13},\,T_{13}^r$ & $\frac{1}{\sqrt{2}}(\ket{1110}-\ket{1011})$\\
\hline 
$T_{14},\,T_{14}^r$ & $\frac{1}{\sqrt{2}}(\ket{1101}-\ket{1011})$\\
\hline 
$T_{23},\,T_{23}^r$ & $\frac{1}{\sqrt{2}}(\ket{1101}-\ket{1011})$ \\
\hline 
$T_{24},\,T_{24}^r$ & $\frac{1}{\sqrt{2}}(\ket{1101}-\ket{0111})$ \\
\hline 
$T_{34},\,T_{34}^r$ & $\frac{1}{\sqrt{2}}(\ket{1011}-\ket{0111})$ \\
\hline
$U_{12/34},\, U_{12/34}^r$  & $\frac{1}{2}(\ket{1111}+\ket{1100}-\ket{0011}-\ket{0000})$ \\
\hline
$U_{13/24},\,U_{13/24}^r$  & $\frac{1}{2}(\ket{1111}+\ket{1010}-\ket{0101}-\ket{0000})$ \\
\hline
$U_{14/23},\,U_{14/23}^r$  & $\frac{1}{2}(\ket{1111}+\ket{1001}-\ket{0110}-\ket{0000})$ \\
\hline 
$V_1,\, V_1^r$  & $\frac{1}{\sqrt{2}}(\ket{1001}-\ket{0111})$\\
\hline 
$V_2,\,V_2^r$  & $\frac{1}{\sqrt{2}}(\ket{1010}-\ket{0111})$\\
\hline 
$V_3,\,V_3^r$  & $\frac{1}{\sqrt{2}}(\ket{1100}-\ket{0111})$ \\
\hline 
$V_4,\,V_4^r$  & $\frac{1}{\sqrt{2}}(\ket{1100}-\ket{1011})$ \\
\hline
$\hat{V}_1^r$ & $\frac{1}{2}(\ket{1100}-\ket{1010}-\ket{0110}-\ket{0000})$ \\
\hline
$\hat{V}_2^r$ & $\frac{1}{2}(\ket{1100}-\ket{1001}-\ket{0101}-\ket{0000})$\\
\hline
$\hat{V}_3^r$ & $\frac{1}{2}(\ket{1010}+\ket{1001}-\ket{0011}-\ket{0000})$\\
\hline
$\hat{V}_4^r$ & $\frac{1}{2}(\ket{0101}+\ket{0101}-\ket{0011}-\ket{0000})$\\
\hline 
$W,\, W^r$ &  $\frac{1}{\sqrt{2}}(\ket{1000}-\ket{0111})$ \\
\hline 
$\hat{W}^r$  & $\frac{1}{2\sqrt{2}}(\ket{1110}+\ket{1101}+\ket{1011}-\ket{1000}  $ \\
                   &$+\ket{0111}-\ket{0100}-\ket{0010}-\ket{0001} )   $ \\
\hline 
$X_{12/34},\, X_{12/34}^r$ & $\frac{1}{2}(\ket{1111}+\ket{1100}-\ket{0011}-\ket{0000})$  \\
\hline 
$X_{13/24},\, X_{13/24}^r$ & $\frac{1}{2}(\ket{1111}-\ket{1010}-\ket{0101}-\ket{0000})$ \\
\hline 
$X_{14/23},\, X_{14/23}^r$ & $\frac{1}{2}(\ket{1111}-\ket{1001}-\ket{0110}-\ket{0000})$  \\
\hline
$\hat{X}_{12}^r$  & $\frac{1}{2}(\ket{1100}-\ket{1011}-\ket{0111}-\ket{0000})$  \\
\hline
$\hat{X}_{34}^r$  &  $\frac{1}{2}(\ket{1110}-\ket{1101}-\ket{0011}-\ket{0000})$  \\
\hline
$\hat{X}_{24}^r$  &  $\frac{1}{2}(\ket{1110}-\ket{1011}-\ket{0101}-\ket{0000})$  \\
\hline
$\hat{X}_{14}^r$  &  $\frac{1}{2}(\ket{1101}-\ket{1011}-\ket{0110}-\ket{0000})$  \\
\hline
$\hat{X}_{13}^r$  &  $\frac{1}{2}(\ket{1101}-\ket{1010}-\ket{0111}-\ket{0000})$ \\
\hline
$\hat{X}_{23}^r$  &  $\frac{1}{2}(\ket{1110}-\ket{1001}-\ket{0111}-\ket{0000})$  \\
\hline
\end{tabular}
\end{center}

\pagebreak

\end{document}